\RequirePackage[2020-02-02]{latexrelease}
\documentclass[showpacs,showkeys,11pt,preprint,preprintnumbers,nofootinbib,
groupedaddress,superscriptaddress,amsmath,amssymb]{revtex4}
\usepackage{xcolor}
\usepackage{amsfonts}
\usepackage{amsmath}
\usepackage{graphicx}
\usepackage{caption}
\usepackage[export]{adjustbox}
\usepackage{verbatim} 
\usepackage{epsfig}
\usepackage{url}
\usepackage{multirow}
\usepackage{hhline}
\usepackage{feynmp}
\usepackage{subfig}

\newcommand{\sqrts}{\mbox{$\sqrt{\mathrm{s}}$}}

\begin{document}

\title{Pair production of heavy charged gauge bosons in $pp$ collisions at LHC}

\author{Ijaz Ahmed}
\email{Ijaz.ahmed@fuuast.edu.pk}
\affiliation{Federal Urdu University of Arts, Science and Technology, Islamabad Pakistan}
\author{Fazal Khaliq}
\email{fazalkhaliq44@gmail.com}
\affiliation{Riphah International University, Hajj Complex, I-14 Islamabad}
\author{M.~U.~Ashraf}
\email{usman.ashraf@cern.ch}
\affiliation{Centre for Cosmology, Particle Physics and Phenomenology (CP3) Université catholique de Louvain, Chemin du Cyclotron, 2, B-1348 Louvain-la-Neuve, Belgium}
\author{Taimoor~Khurshid}
\email{taimoor.khurshid@iiu.edu.pk}
\affiliation{International Islamic University, H-10, Islamabad Pakistan}
\author{Jamil~Muhammad}
\email{mjamil@konkuk.ac.kr}
\affiliation{Sang-Ho College, and Department of Physics, Konkuk University, Seoul 05029, South Korea}

\begin{abstract}
Two oppositely charged new heavy gauge boson pair production at the Large Hadron Collider (LHC), is presented in this paper. These bosons are known as $W^{'}$ boson due to the reason that it is the heavy version of Standard Model's weak force carrier, the $W$ boson. The production cross section and decay width in proton-proton ($pp$) collision at \sqrts~= 8 TeV are calculated for different masses and coupling strengths of $W^{'}$. Efficiencies for different signal regions and branching ratios for different decay channels are computed. In this study, the pair production ($W^{'^{+}}W^{'^{-}}$) is considered in emerging new physics as a result of $pp$ collision at \sqrts~= 8 TeV at the LHC with final state containing two tau ($\tau$) leptons and two neutrinos (each $W^{'}$ decay to $\tau$ and its neutrino). The event selection efficiency similar to the CMS experiment is used for the mass of $W^{'}$ to set lower limits for different coupling strengths of $W^{'}$ and, the obtained results are presented in this work. 
For heavy gauge bosons, when coupling strength is similar to that of Standard Model's $W$ boson, the mass of $W^{'}$ below 445 GeV are excluded at a confidence level of $95\%$.

\end{abstract}
\keywords{heavy gauge boson, LHC, tau leptons, proton Collider}

\pacs{12.60}

\maketitle

\section{Introduction}
A new physics can be observed at TeV energy scale. The new scenario of physics is the finding of additional new heavy gauge bosons ($W^{'^{\pm}}, Z^{'}$). Many extensions of the standard model realize the existence of these additional gauge bosons. These bosons are the heavy version of weak vector gauge bosons of the standard model (SM). The properties of these bosons may be similar or not to that of the standard model weak bosons $W$ which depends upon the underlying theory. The model that predict heavy $W^{'}$ bosons also contains $Z^{'}$ bosons generically, but this is not true in the reverse~\cite{lab1}. 

Different theories and models predict the existence of heavy charged gauge bosons(W'). The standard model W bosons, in theories with extra dimension, may propagate in extra dimension, and gives rise to heavy copies [2].  One approach is the little higgs models. These models predict the existence of W' bosons, with the idea that the higgs boson is a nambu-goldstone boson [3]. In Left Right Symmetric Model (LRSM) charged gauge boson may be realized in symmetric way that can be left, right or both-handed [4, 5, 6]

The detail of the model gives the difference in mass of $W^{'}$ and $Z^{'}$ bosons, hence the discovery of $W^{'}$ bosons is more probable than the discovery of $Z^{'}$ bosons. The property that differentiates standard model $W$ and the new heavy charged gauge bosons is that it may couple to left-handed, right-handed or mixture of both fermions while standard model weak bosons only couple to left-handed fermions. The lagrangian that generally gives mathematical description of fermions interaction of $W^{'}$ bosons is described in~\cite{lab1}.

\begin{equation} 
\label{eq1}
L=\frac{V_{ij}}{2\sqrt{2}} \bar{f_i}{\gamma }_\mu ( g'_{R}(1+ \gamma^5 ) + g'_{L} (1 - \gamma^{5} ) ) {W'}^\mu f_j + h.c 
\end{equation} 

Where left-hand (right-hand) coupling constant is given by $g^{'}_{L(R)}$ respectively. The $V_{i,j}$ is the $3 \times 3$ identity matrix for lepton and it represents CKM matrix for quarks. The left and right-handed chiral projection operators are given by $(1 \pm \gamma^{5})$. In case when $g^{'}_{R} = 0$ and $g^{'}_{L}\neq 0$, then $W^{'}$ boson has purely left-hand coupling where it couple with both leptons and quarks, while when $g^{'}_{R}\neq 0$ and $g^{'}_{L} = 0$, then only $W^{'}$ boson wil have purely right-handed coupling but only with quarks. Since the right-handed neutrinos do not exist, so coupling of $W^{'}$ boson is restricted with the leptons including neutrinos or the neutrino mass must be much higher than $W^{'}$ bosons, which is also not possible as the neutrinos masses are confirmed lying at few electron volts.

\section{Search of heavy gauge bosons ($W^{'}$)}
\begin{figure}
\begin{center}
\includegraphics[width=6cm]{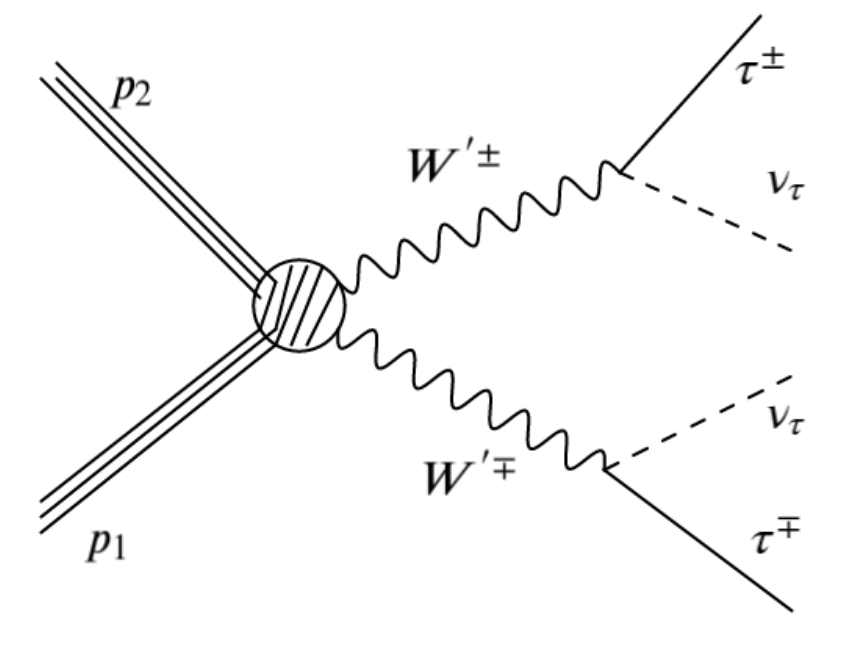}
\caption{The pair of heavy charged gauge bosons decayed to $\tau$ and its neutrino in $pp$ collisions }
\label{fig1}
\end{center}
\end{figure}

Different signatures are used in many experiments to search for $W^{'}$ bosons. In ATLAS experiment $W^{'}$ bosons are considered to decay into lepton along with missing transverse energy from neutrinos~\cite{lab2}, where it excludes $W^{'}$) masses less than 5.1 TeV in standard model conditions at $95\%$ confidence level. This paper excludes two main searches that can be done for the heavy-charged gauge boson via direct and Indirect searches.\\
In the past, extensive work on heavy charged gauge boson is performed both phenomenologically as well as experimentally e.g., in Ref.~\cite{atlas} ATLAS collaboration studied single $W^{'}$ production at LHC $pp \rightarrow W^{'}\rightarrow l\nu$ using 139 $fb^{-1}$ data at $\sqrt{s}$ = 13 TeV where $(l=e$ or $l=\mu)$ are extracted in the model-independent searches. Similarly, CMS experiment~\cite{cms} explored $pp \rightarrow W^{'}\rightarrow tb$ channel, where an integrated luminosity 2.6 $fb^-1$ at $\sqrt{s}$ = 13 TeV is used and results provide the most stringent limits for right-handed $W^{'}$ bosons in the top and bottom quark decay channel. A few research articles~\cite{anatomy, china, iran} are also reported in which analysis is carried out either at $\sqrt{s}$ = 8 TeV in $\tau\nu$ final state or at $\sqrt{s}$ = 14 TeV in High Luminosity LHC (HL-LHC).\\
In current study, two oppositely charged heavy gauge ($W^{'}$) bosons are produced in $pp$ collisions at $\sqrt{s}$ = 8 TeV as shown in figure~\ref{fig1}. Since this energy is easily accessible at LHC and abundant of such particles are produced where it can decays into $\tau$ leptons and its neutrinos final state (each $W^{'}$ decays to one $\tau$ and its neutrino). Due to the presence of neutrino in the final state $g^{'}_{L}\neq 0$. The efficiencies calculated by CMS collaboration~\cite{lab3} are used in our study to compare the required signal yields having standard model backgrounds. This makes us able to set lower limits on $W^{'}$ mass. The integrated luminosity used is 18.1 $fb^{-1}$ and 19.6 $fb^-{1}$ for two different channels.\\

Many experimental analyses propose left or right-handed $W^{'}$ in the direct searches for $W^{'}$ bosons production at hadron colliders. These bosons decay into leptons in the final state with standard model-like couplings. The decay of right-handed or left-handed bosons into right-handed neutrino is kinematically restricted to the conditions on mass of $W^{'}$ that is $m_{W^{'}} > 786$ GeV \cite{lab4,lab5,lab6,lab7,lab8,lab9,lab10,lab11}. This decay is forbidden if mass of the right-handed neutrino is greater than mass of $W^{'}$ bosons. In di-jet data, right-handed $W^{'}$ bosons are directly restricted by peak search only. The light quarks have larger coupling with $W^{'}$ until to reach the limit for mass of 420 GeV \cite{lab12,lab13,lab14} as limited by the di-jet QCD background. \\ 

\begin{figure}
\begin{center}
\includegraphics[width=6cm]{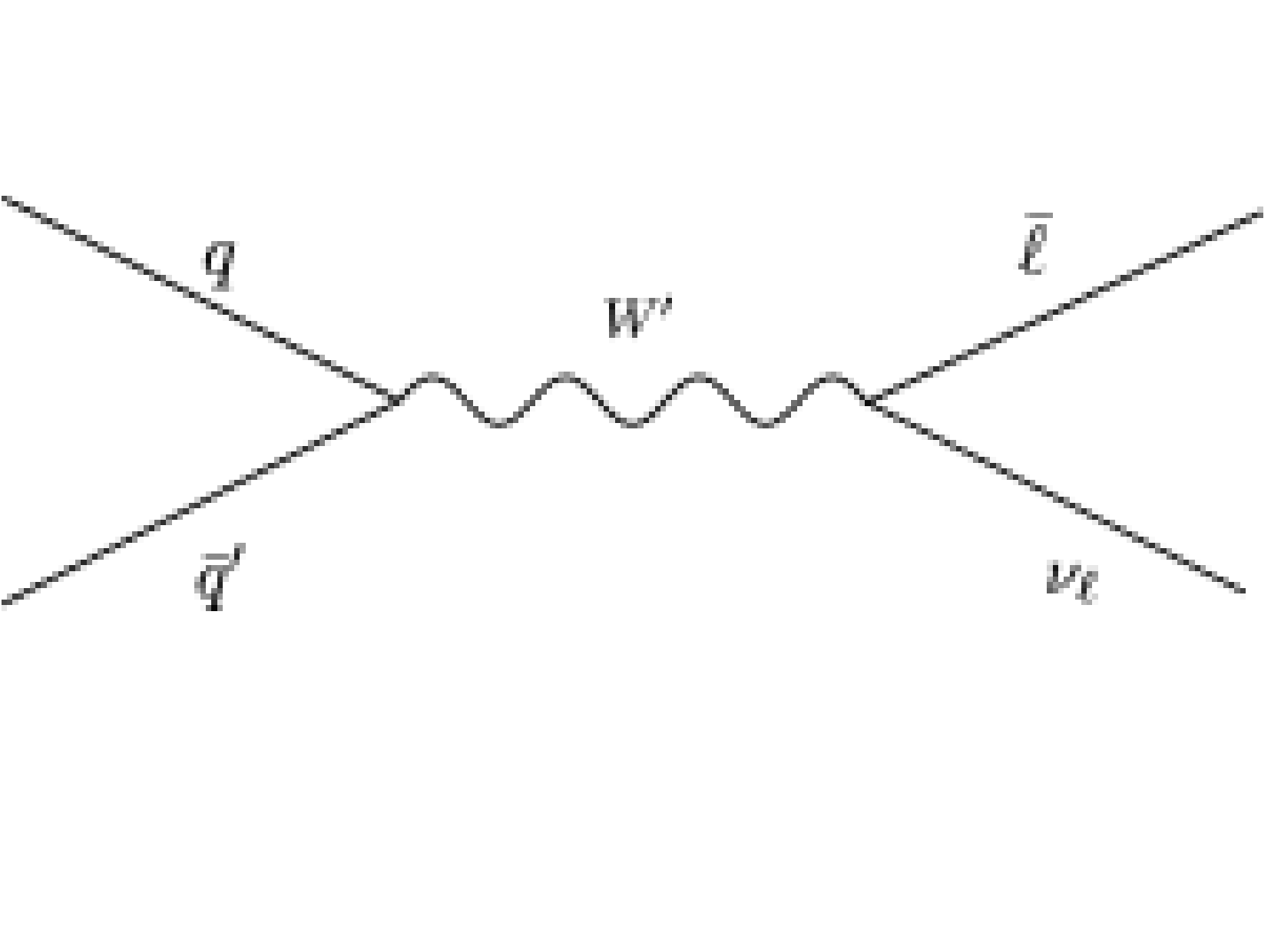}
\caption{ $W'$ production via quark anti-quark annihilation, and decays to lepton and its corresponding neutrino}
\label{fig2}
\end{center}
\end{figure}
$W'$ boson can be detected directly at the LHC, by decaying to lepton and its neutrino or top and bottom quark in quark anti-quark annihilation is shown in figure~\ref{fig2}. This physics can be achieved at the energy scale of TeV.   

\subsection{Alternative search}  
\begin{figure}
\begin{center}
\includegraphics[width=5cm]{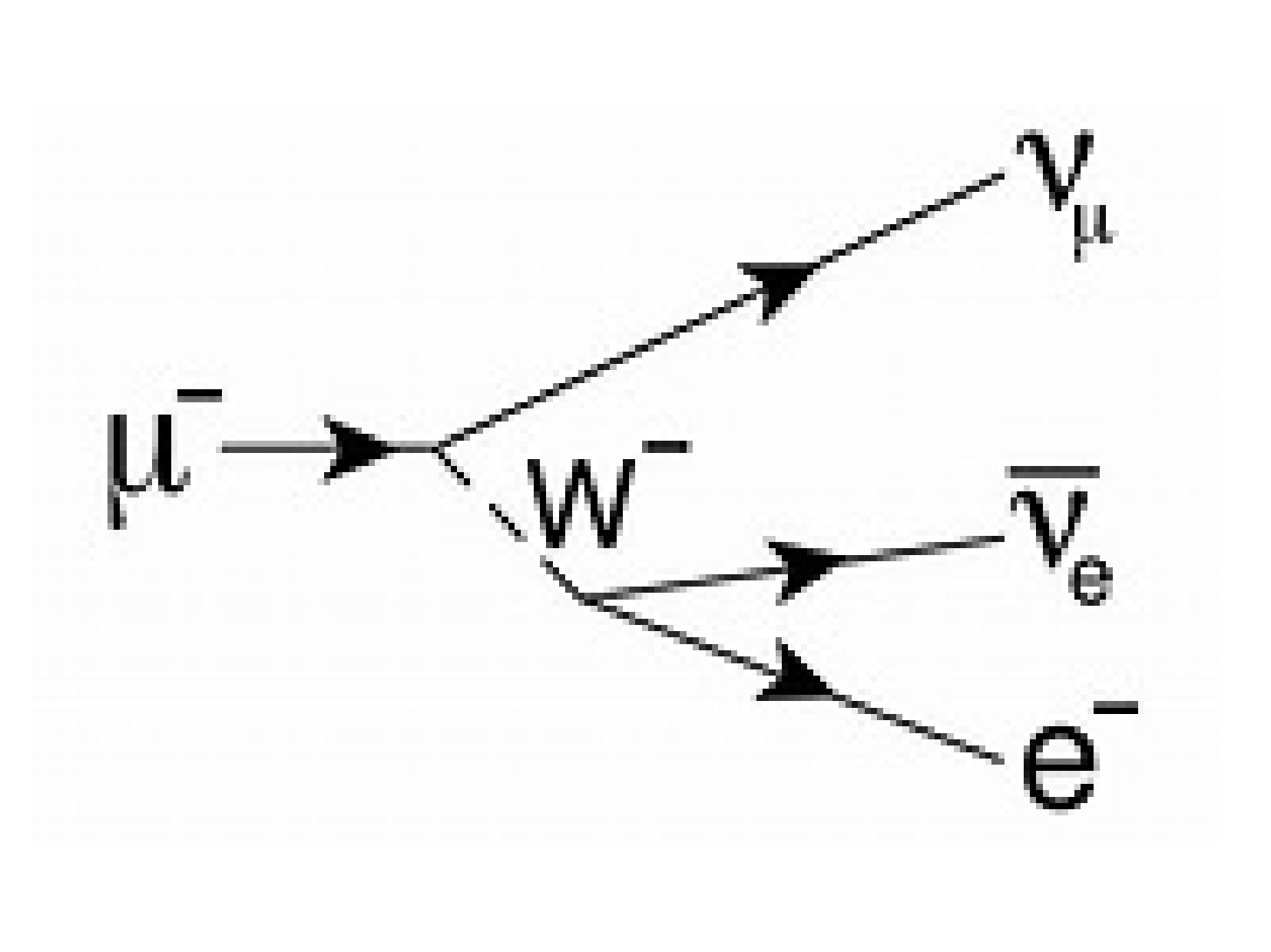}
\caption{A view of muon decay to electron and neutrinos }
\label{fig3}
\end{center}
\end{figure}

The second method for the search of $W^{'}$ is the indirect method. The Standard Model $W$ can be replaced by $W^{'}$ in some decay process like muon decay to set a limit on its mass for study. Figure~\ref{fig3} presents a schemetic view of muon decay in which the SM $W$ can be replaced in this process with the heavy charged gauge boson.\\
\section{Simulation and Results}
\subsection{Cross section and decay width calculation}
For the theoretical study of pair production of $W^{'}$ at \sqrts~= 8 TeV in $pp$ collision and its decay into $\tau$ and its neutrino in the final state, Madgraph 2.6.0 \cite{lab15} is used. It is the extension to Madgraph5 which is matrix-element generator.
The events are generated in Madgraph, first considering only left-handed $W^{'}$ by setting the coupling parameter for $W^{'}$($g^{'}_{L}$ = $g_{SM}$ , $g^{'}_{R}$ = 0) the interaction as permitted both to quarks and leptons. The mass of $W^{'}$ boson is varied from 110 GeV to 500 GeV with an increment of 30 GeV, selecting the total energy of beam 14 TeV (7 TeV for each beam of proton).\\
The decay width or lifetime was calculated using Madgraph for various masses and compared with the results in \cite{lab1}. In this case the partial width for $W^{'}$ decays to $t\bar{b}$ and $\bar{t}b$ and the total decay width for $W^{'}$ in leading order and next to leading order precision is calculated. The results are in good agreement. The decay width is the function of modes of decay, process of decay, coupling constant which depend upon kinematics constraint. Estimation for production cross-section are also done by MadGraph. Decay width and production cross section for both quarks and leptons are given in table \ref{tab1}. The model does not keep the value of coupling constant fixed.\\
Then we modified the coupling constant to find its effect on the production cross section and decay width, by increasing and decreasing (multiplying the left-hand coupling with 1.5 and 0.5 of the standard model) coupling constant. By this increase or decrease in coupling constant, cross-section also increases by factor of 5 and decreases by factor 0.062. The decay width increases or decreases by a factor of 2.25 and 0.25 respectively.\\
Now the left-handed and right-handed couplings are changed such that the sums of their square are equal to the square of standard model coupling.
\begin{equation}
\label{eq2}
g_{SM}^{2} = (g^{'}_{L})^{2} + (g^{'}_{R})^{2}
\end{equation}
and
\begin{equation}
\label{eq3}
g^{'}_L = g_{SM}\cos{\theta}
\end{equation}
Where $\theta$ is the mixing angle, from the above equation it is observed that by changing angle from $0^0$ to $90^0$, $W^{'}$ goes to purely right-handed from purely left-handed. The cross-section and decay width with different mixing angles for mass of $W^{'} = 350$ GeV are given in Table \ref{tab2}.

\begin{table}[ht]
\begin{center}
\resizebox{\textwidth}{!}{%
\begin{tabular}{|c|c|c|c|c|c|c|c|} \hline
\multirow{2}{*}{\begin{tabular}[c]{@{}l@{}} Mass \\ (GeV)\end{tabular}}  &  \multicolumn{2}{|c|}{$g^{'}_{R} = 0$, $g^{'}_{L} = 0.32$} & \multicolumn{2}{|c|}{$g^{'}_{R} = 0$, $g^{'}_{L} = 0.64$} & \multicolumn{2}{|c|}{$g^{'}_{R} = 0$, $g^{'}_{L} = 0.96$}  \\  & $\sigma(pp\rightarrow W^{'^{+}}W^{'^{-}}$) (fb) & $\Gamma(W^{'} \rightarrow X Y$) (GeV) & $\sigma(pp\rightarrow W^{'^{+}}W^{'^{-}}$ & $\Gamma(W^{'} \rightarrow X Y$) & $\sigma(pp\rightarrow W^{'^{+}}W^{'^{-}}  (fb)$ & $\Gamma(W^{'} \rightarrow X Y$) (GeV) \\ \hline
 110 & 128.6 & 0.69 &2055.3 & 2.8 & 10296.9 & 6.2 \\ \hline
 140 & 57.3&0.82&923.5&3.5&3950.8&7.9 \\ \hline
 170 & 22.4&1.1&360.1&4.3 &1804.5 &9.5 \\ \hline
 200 & 10.9 &1.3 &173.9 &5.2 &867.2 &11.7 \\ \hline
 230 & 5.5 &1.6 &87.6 &6.2 &440 &14 \\ \hline
 260 & 3.0 &1.8 &48 &7.4 & 238 &16.6 \\ \hline
 290 & 1.7& 2.1& 28.1& 8.5& 146.5& 19.1 \\ \hline
 320 & 1.1& 2.4& 17.4& 9.6& 84.5& 21.6  \\ \hline
 350 & 0.7 &2.7& 11.2& 10.7& 40.2& 24.0 \\ \hline
 380 &0.5& 2.9& 7.5& 11.7& 19.2& 24.4 \\ \hline
 410 &0.3 &3.20& 5.16& 12.82& 9.2& 28.8 \\ \hline
 440 &0.2 &3.47& 3.64& 13.88& 4.6& 31.2 \\ \hline
 470 &0.2& 3.7 &2.6& 14.9& 2.9& 33.6 \\ \hline
 500 &0.12& 4.0& 1.9& 15.9& 1.2& 35.9 \\ \hline 
 \end{tabular}
}
\caption{Cross section and decay width for different masses and different coupling constants.} 
\label{tab1}
\end{center}
\end{table}

\begin{table}[ht]
\begin{center}
\begin{tabular}{|c|c|c|c|c|} \hline
Mixing angle & Coupling constant & Cross section (fb)&Decay Width (GeV)& $\tau^{+}\nu_{e}$  \\ \hline
0& $g^{'}_{R}=0, g^{'}_{L}=0.64$ & 11.2&10.7& 0.09 \\ \hline
15 & $g^{'}_{R}=0.16, g^{'}_{L}=0.62$ & 9.97& 10.7&0.09 \\ \hline
30 & $g^{'}_{R}=0.32, g^{'}_{L}=0.56$ & 7.3&10.9&0.09 \\ \hline
45 & $g^{'}_{R}=0.45, g^{'}_{L}=0.45$ & 5.7&10.6&0.09 \\ \hline
60 & $g^{'}_{R}=0.56, g^{'}_{L}=0.32$ & 7.8 &10.9&0.09 \\ \hline
90 & $g^{'}_{R}=0.64, g^{'}_{L}=0$ & 11.3 & 10.7& 0 \\ \hline
\end{tabular}

\caption{The cross section and decay width for $m_{W}^{'}$ = 350 GeV for different mixing angles and branching ratios in $\tau \nu_{\tau}$ final state} 
\label{tab2} 
\end{center}
\end{table}

\subsection{Branching ratios} 
\begin{figure}
\begin{center}
\includegraphics[width=10cm]{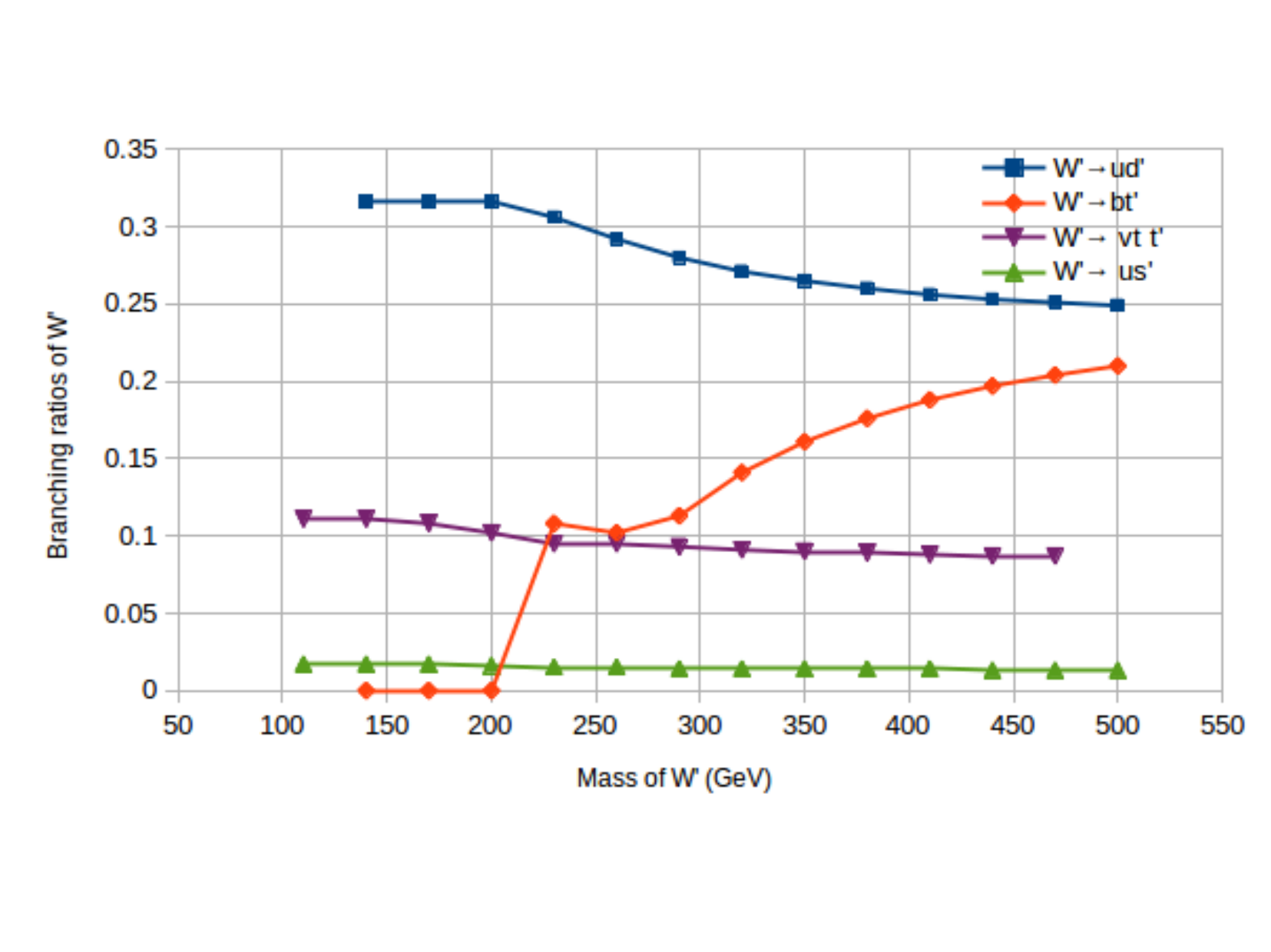}
\caption{ Branching ratios of $W^{'}$ decays}
\label{fig4}
\end{center}
\end{figure}

$W^{'}$ boson decay to different final states, in the current study we only consider $\tau$ and its neutrino in the final state. TAUOLA package \cite{lab17} is used for the simulation of $\tau$ lepton decays. This package is also used for the leptonic and hadronic decay of taus, to simulate in the final state. It also gives full information of neutrinos and mediator particles in the final state. It also contains spin information and can do simulations for the angular distribution of decay products. The ratio of decay in one channel divided by the total decay width is referring as the branching ratio. This is given by TAUOLA package and listed in table \ref{tab3} and is shown in fig. \ref{fig4} for different masses and different decay channels.
All the signal processes are produced with MadGraph5 2.3.3 [24]. The output of both these packages in Les Houches Event Format (LHEF) is used by PYTHIA 8.1.5.3 \cite{lab30} for partonic showering, gluon radiation, fragmentation and hadronization. Taoula package \cite{lab17} used for two Tau decay process in simulation. The Tau decay width can be separated due to its narrow decay width. This package simulates both the decay of Tau in leptonic and hadronic decay. 

\begin{table}[ht]
\begin{center}
\begin{tabular}{|c|c|c|c|c|c|c|c|c|} \hline
 Mass  & \multicolumn{8}{c|}{ Branching ratios ($W^{'} \rightarrow x + y$) }\\ 
(GeV) & $u+d'$ & $c+s'$ & $b+t'$ & $e + e$ & $\mu + \mu $ &$\tau + \tau'$ & $u+s' $ & $c+d'$ \\ \hline
110 & 0.32 & 0.32 & 0 & 0.11 & 0.11 & 0.11 & 0.02 &	0.02 \\ \hline
140 & 0.32 & 0.32 & 0 & 0.11 & 0.11 & 0.11 & 0.02 & 0.02 \\ \hline
170 & 0.32 & 0.32 & 0 & 0.11 & 0.11 & 0.11 & 0.02 & 0.02 \\ \hline
200 & 0.31 & 0.31 & 0.03 & 0.11 & 0.108 & 0.11 & 0.02 & 0.02 \\ \hline
230 & 0.29 & 0.29 & 0.08 & 0.10 & 0.10 & 0.10 & 0.02 & 0.02 \\ \hline
260 & 0.28 & 0.28 & 0.11 & 0.10 & 0.10 & 	0.10 & 0.02 & 0.01 \\ \hline
290 & 0.27 & 0.27 & 0.14 & 0.10 & 0.10 & 0.10 & 0.01 & 0.01 \\ \hline
320 & 0.26 & 0.26 & 0.16 & 0.09 & 0.09 & 0.09 & 0.01 & 0.01 \\ \hline
350 & 0.26 & 0.26 & 0.18 & 0.09 & 0.09 & 0.09 & 0.01 & 0.01 \\ \hline
380 & 0.26 & 0.26 & 0.19 & 0.09 & 0.09 & 0.09 & 0.01 & 0.01 \\ \hline
410 & 0.25 & 0.25 & 0.11 & 0.09 & 0.09 & 0.09 & 0.01 & 0.01\\ \hline
440 & 0.25 & 0.25 & 0.20 & 0.09 & 0.09 & 0.09 & 0.01 & 0.01\\ \hline
470 & 0.25 & 0.25 & 0.21 & 0.09 & 0.09 & 0.09 & 0.01 & 0.01\\ \hline
500 & 0.24 & 0.25 & 0.22 & 0.09 & 0.09 & 0.09 & 0.01 & 0.01\\ \hline
\end{tabular}
 \caption{  Branching ratios of $W^{'}$ for different signal channels.}
 \label{tab3}
\end{center}
\end{table}
\subsection{Selection efficiency}

The events are generated by MadGraph and $\tau$ decay is performed by Tauola package. Now we will obtain the selection efficiencies for different masses of $W^{'}$ at different coupling constants. For signal selection efficiency cuts as reference \cite{lab2} have been used to find the selection cuts for the probability to pass for any given signal region. Efficiency cuts are reported in this paper for events reconstructed properties and these cuts are the functions of generator-level values for that property. The detector's effects are taken into account. All efficiency cuts are multiplied for different signals regions and different channels to get full selection efficiency. 


\begin{table}[ht]
\begin{center}
\begin{tabular}{|c|c|c|c|c|} \hline
Mass (GeV) & SR-1 & SR-2 & $\mu \tau $ & $e \tau$ \\ \hline
110 & 0.09 & 0.77 & 0.12 & 0.01 \\ \hline
140 & 0.27 & 0.92 & 0.40 & 0.03 \\ \hline
170 & 0.09 & 0.77 & 0.01 & 0.01 \\ \hline
200 & 0.55 & 1.00 & 0.14 & 0.10 \\ \hline
230 & 1.20 & 1.10 & 0.48 & 0.39 \\ \hline
260 & 1.51 & 1.05 & 0.68 & 0.54 \\ \hline
290 & 1.85 & 1.05 & 0.90 & 0.72 \\ \hline
320 & 2.16 & 1.07 & 1.08 & 0.92 \\ \hline
350 & 2.46 & 0.99 & 1.37 & 1.10 \\ \hline
380 & 2.85 & 0.98 & 1.56 & 1.30 \\ \hline
410 & 3.00 & 1.01 & 1.92 & 1.57 \\ \hline
440 & 3.29 & 0.96 & 2.15 & 1.73 \\ \hline
470 & 3.40 & 0.96 & 2.42 & 2.05 \\ \hline
500 & 3.70 & 0.83 & 2.63 & 2.14 \\ \hline
\end{tabular}
\caption{ Efficiencies of different channels for different masses in the standard model like scenario}
 \label{tab4}
\end{center}
\end{table}

\begin{table}[ht]
\begin{center}
\begin{tabular}{|c|c|c|} \hline
 Channel & Integrated luminosity($fb^-1$) & Uncertainty (\%) \\ \hline
 $\tau_h$ $\tau_h$ & 18.1 & 20 \\ \hline
 Lepton $\tau_h$ & 19.6 & 25 \\ \hline
\end{tabular}
\caption{ Integrated luminosity and uncertainty of the channels}
 \label{tab5}
\end{center} 
\end{table}
 
\begin{figure}
\begin{center}
\includegraphics[width=8.5cm]{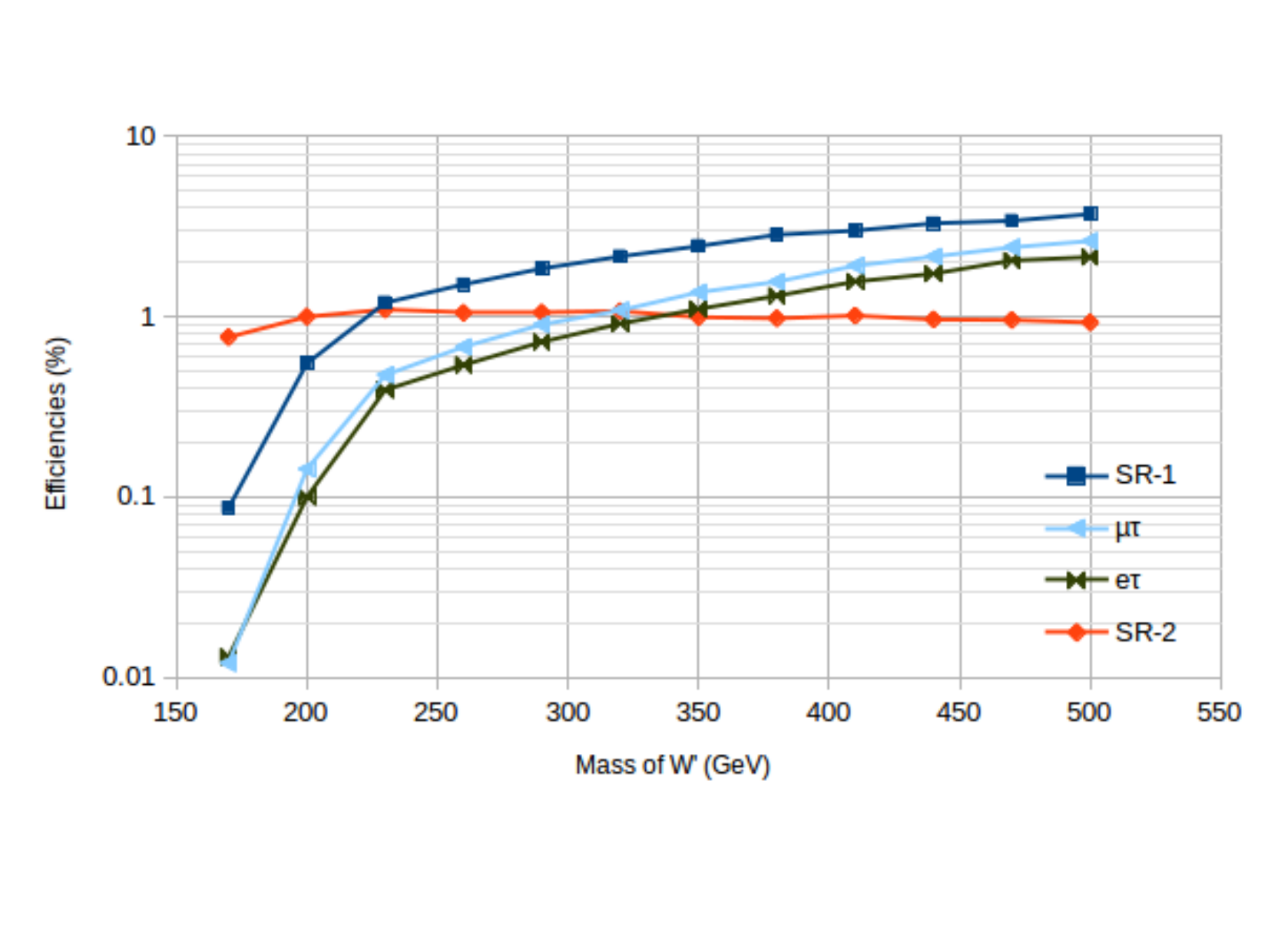}
\caption{ Mass verses efficiency for different channels in the standard model like scenario}
\label{fig5}
\end{center}
\end{figure}

The efficiency for different channels are calculated for different masses of $W^{'}$ bosons by running simulation code. The table \ref{tab4} gives the efficiency for different channels and different masses in the standard model like scenario as shown in fig. \ref{fig5}.
The efficiencies are calculated for different masses of $W^{'}$ and for different coupling constants (mixing angles). These results are compared with the given table \ref{tab4} and fig. \ref{fig5} is produced in standard model like scenario, and found that the efficiency do not depend upon the coupling strength as expected. The efficiency of the signal region is only the function of kinematics of the event generated which changes with mass of $W^{'}$.\\

For any given integrated luminosity, the total number of expected events can be estimated in that channel, if one knows the production cross-section of the event and branching ratio (BR) of the signal and for that channel full selection efficiency by the following equation.
\begin{equation}
\label{eq4}
N = \mathcal{L} \times \sigma(pp \rightarrow W^{'+} W^{'-} ) \times BR(W^{'} \rightarrow \tau \nu) \times BR(W^{'} \rightarrow \tau \nu) \times \epsilon^{ch}
\end{equation}
Where $\epsilon^{ch} $ is the full selection efficiency. The systematic uncertainty and integrated luminosity are given in the table \ref{tab5} taken from the experimental paper followed.

\begin{table}[ht]
\begin{center}
\begin{tabular}{|c|c|c|c|c|c|c|c|c|c|c|} \hline
Mass (GeV) & $\sigma$  (fb) & \multicolumn{2}{|c|}{Luminosity} & \multicolumn{2}{|c|}{Luminosity} & $\Gamma$ & \multicolumn{4}{|c|}{Expected number of Events in different signal regions} \\ 
&  & \multicolumn{2}{|c|}{$\mathcal{L} = 18.1fb^{-1}$} & \multicolumn{2}{|c|}{$\mathcal{L} = 19.6 fb^{-1}$} & & \multicolumn{4}{|c|}{}   \\ \hline
&  & SR-1 & SR-2 & $\mu\ \tau$ & $e\ \tau$ &  & SR-1 & SR-2 & $\mu\ \tau$ & $e\ \tau$ \\ \hline
110 & 2055.28 & 0.06 & 0.52 & 0.01 & 0.01 & 0.11 & 27 & 234 & 4.8 & 4.8 \\ \hline
140 & 923.55 & 0.20 & 0.64 & 0.03 & 0.03 & 0.11 & 40.45 & 129.45 & 6.57 & 6.57 \\ \hline
170 & 360.11 & 0.46 & 0.76 & 0.11 & 0.09 & 0.11 & 36.28 & 59.93 & 9.40 & 7.68 \\ \hline
200 & 173.94 & 0.75 & 0.81 & 0.24 & 0.12 & 0.10 & 23.61 & 25.50 & 8.18 & 4.09 \\ \hline
230 & 87.60 & 1.06 & 0..89 & 0.38 & 0.31 & 0.10 & 16.80 & 14.11 & 6.52 & 5.32 \\ \hline
260 & 48.09 & 1.41 & 0.89 & 0.62 & 0.48 & 0.10 & 12.40 & 7.82 & 5.90 & 4.57 \\ \hline
290 & 28.08 & 1.67 & 0.88 & 0.98 & 0.70 & 0.09 & 6.87 & 3.62 & 4.37 & 3.12 \\ \hline
320 & 17.41 & 1.98 & 0.90 & 1.06 & 0.82 & 0.09 & 5.05 & 2.30 & 2.92 & 2.26 \\ \hline
350 & 11.21 & 2.29 & 0.93 & 1.32 & 1.11 & 0.09 & 3.76 & 1.53 & 2.35 & 1.97 \\ \hline
380 & 7.50 & 2.60 & 0.94 & 1.53 & 1.21 & 0.09 & 2.86 & 1.03 & 1.82 & 1.44 \\ \hline
410 & 5.16 & 2.92 & 0.88 & 1.76 & 1.50 & 0.09 & 2.21 & 0.66 & 0.15 & 0.14 \\ \hline
440 & 3.64 & 3.13 & 0.91 & 2.03 & 1.69 & 0.09 & 1.67 & 0.48 & 1.17 & 0.97 \\ \hline
470 & 2.61 & 3.36 & 0.92 & 2.28 & 1.92 & 0.09 & 1.28 & 0.35 & 0,94 & 0.79 \\ \hline
500 & 1.91 & 3.57 & 0.84 & 2.54 & 2.19 & 0.09 & 1.00 & 0.24 & 0.77 & 0.66 \\ \hline
\end{tabular}
\caption{ The expected number of Events in Standard Model scenario}	
\label{tab6}
\end{center}
\end{table}

	The number of expected events are calculated using above formula for production cross section for the standard model scenario and branching ratios of the signal of our interest ($W' \rightarrow \tau \nu$) using luminosity of the above table taken from experimental paper and the channel efficiencies listed in table \ref{tab6} are calculated using our simulation codes. 

\subsection{Transverse mass }
	The detector can detect the transverse component indirectly, although detector cannot directly detect the emerging neutrinos. According to the momentum conservation the final state should not have any transverse component, because the initial transferable components of momentum are zero. When all transverse components are added and their sum is other than zero, the additional transverse component to make zero is known as missing transverse energy (MET). This energy represents neutrinos which are the only particle in the Standard Model which contribute to the missing energy. The signal missing transverse energy of the event is not because of instrumental MET but related to the real physical contents. If we do not have the invariant mass, the transverse component may be reconstructed for neutrino. {This is known to be  called as transverse mass ($M_T$) rather be calculated as:

\begin{equation}
\label{eq5}
 M_T = \sqrt{2P^{\tau}_T P^{\nu}_T (1 - \cos{\Delta {\theta}_{\tau ,\nu}} )} 
 \end{equation} 

Equation~\ref{eq5} represents the transverse momentum of $\tau$, the missing transverse energy which is the transverse component of neutrino.

For the generated events, this was done as a cross check, and kinematics are produced. In the final state of our signal contain mixture of hadronic, leptonic and also pure hadronic channels. For different mass of $W^{'}$ the distribution of missing transverse momentum ($P_{T}^{miss}$) and transverse mass $M_{T2}$ are given in figs. \ref{fig6}(a), \ref{fig6}(b), \ref{fig7}(a) and \ref{fig7}(b) for different channels. The distribution of transverse momentum are given in fig. \ref{fig8}(a) and \ref{fig8}(b) of the $\tau_h$ lepton in $l\tau_h$ channels. Looking into the plot of the distribution of these variables, it is observed that heavier  objects are produced with increasing mass of $W^{'}$ bosons. The transverse gives Jacobeans peak characteristics instead of Breit Weigner which in case of invariant mass the peak rise upto $M_{T} = M_{W^{'}}$ with transverse mass and then start falling rapidly. For the statistical analysis transverse distribution is very useful.

\begin{figure}[!tbp]
\centering
\subfloat[]{\includegraphics[width=8cm]{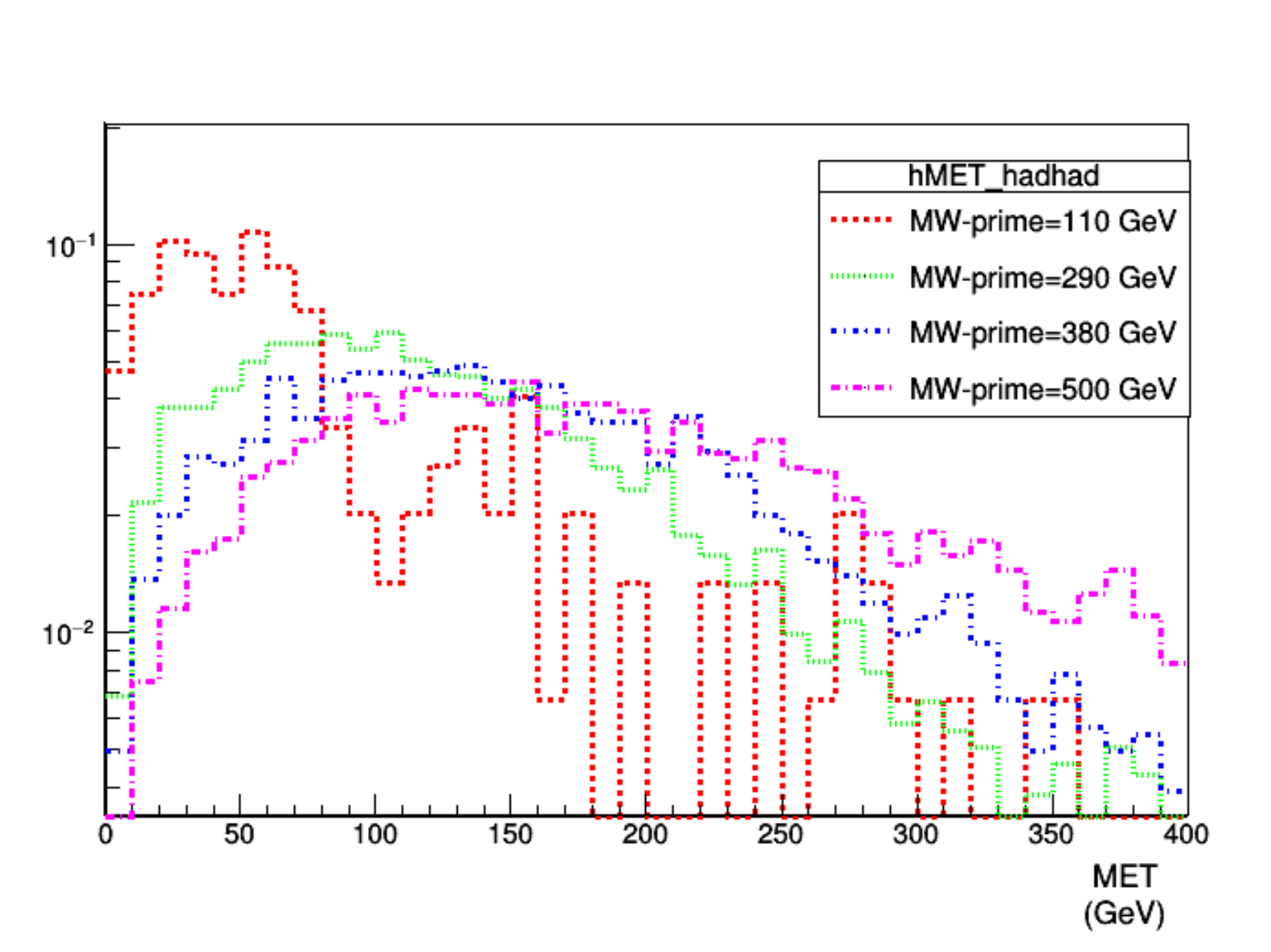} }%
\qquad
\subfloat[]{\includegraphics[width=8cm]{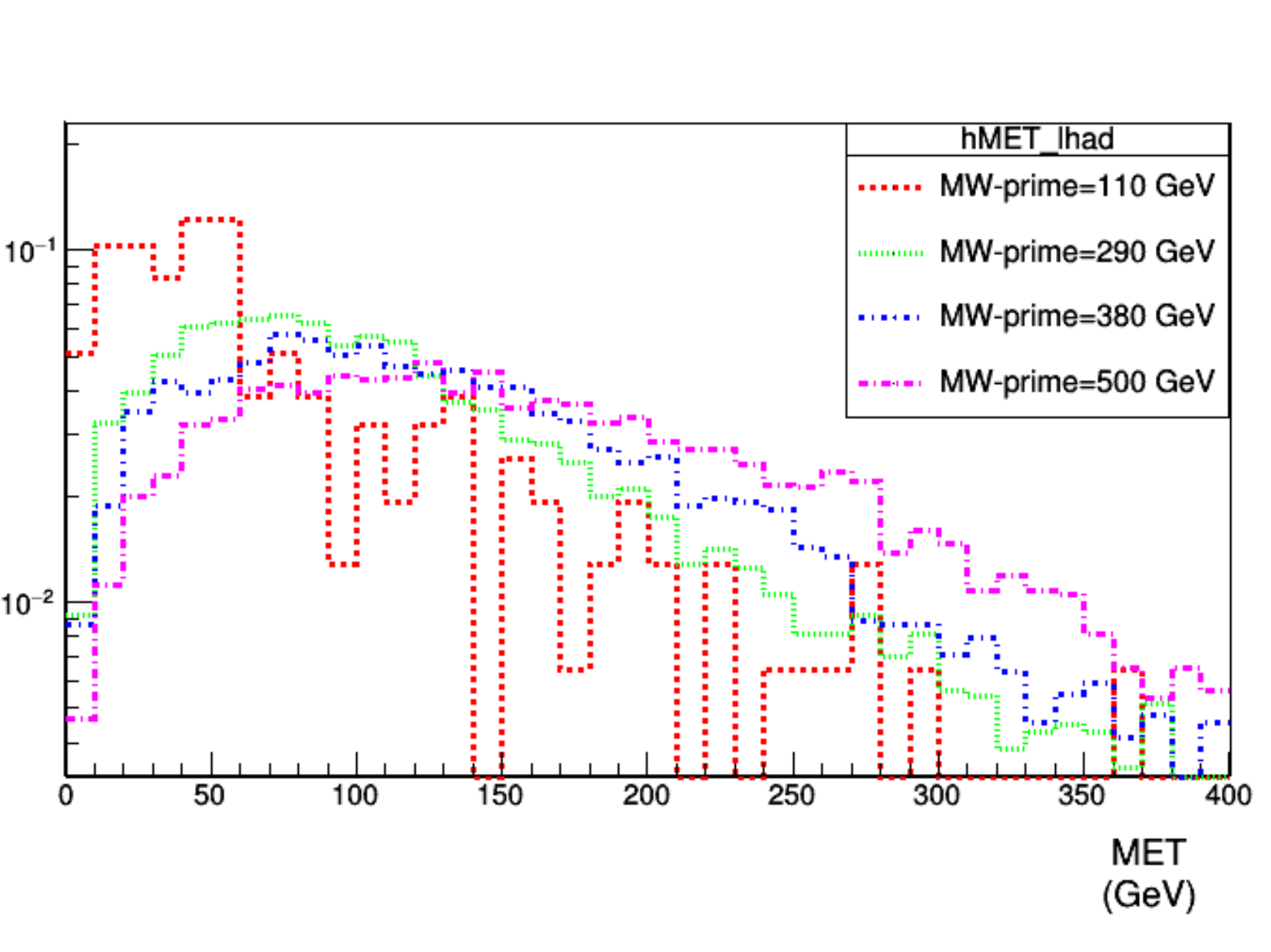} }%
\caption{The missing transverse energy distribution is shown for a) fully hadronic and b) semi-leptnic channel}%
\label{fig6}\quad 

\end{figure}

\begin{figure}%
\centering
\subfloat[]{{\includegraphics[width=8cm]{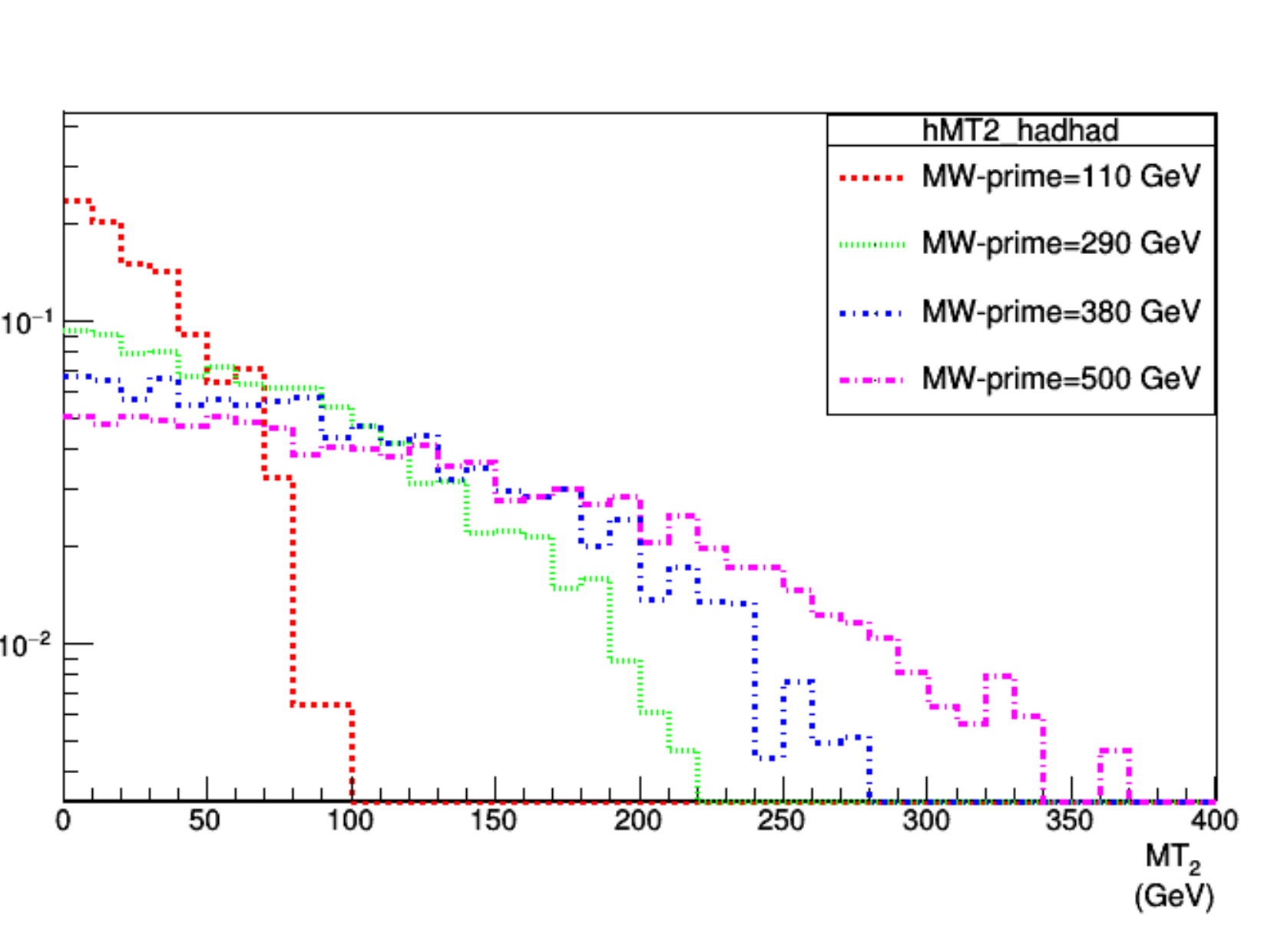} }}%
\qquad
\subfloat[]{{\includegraphics[width=8cm]{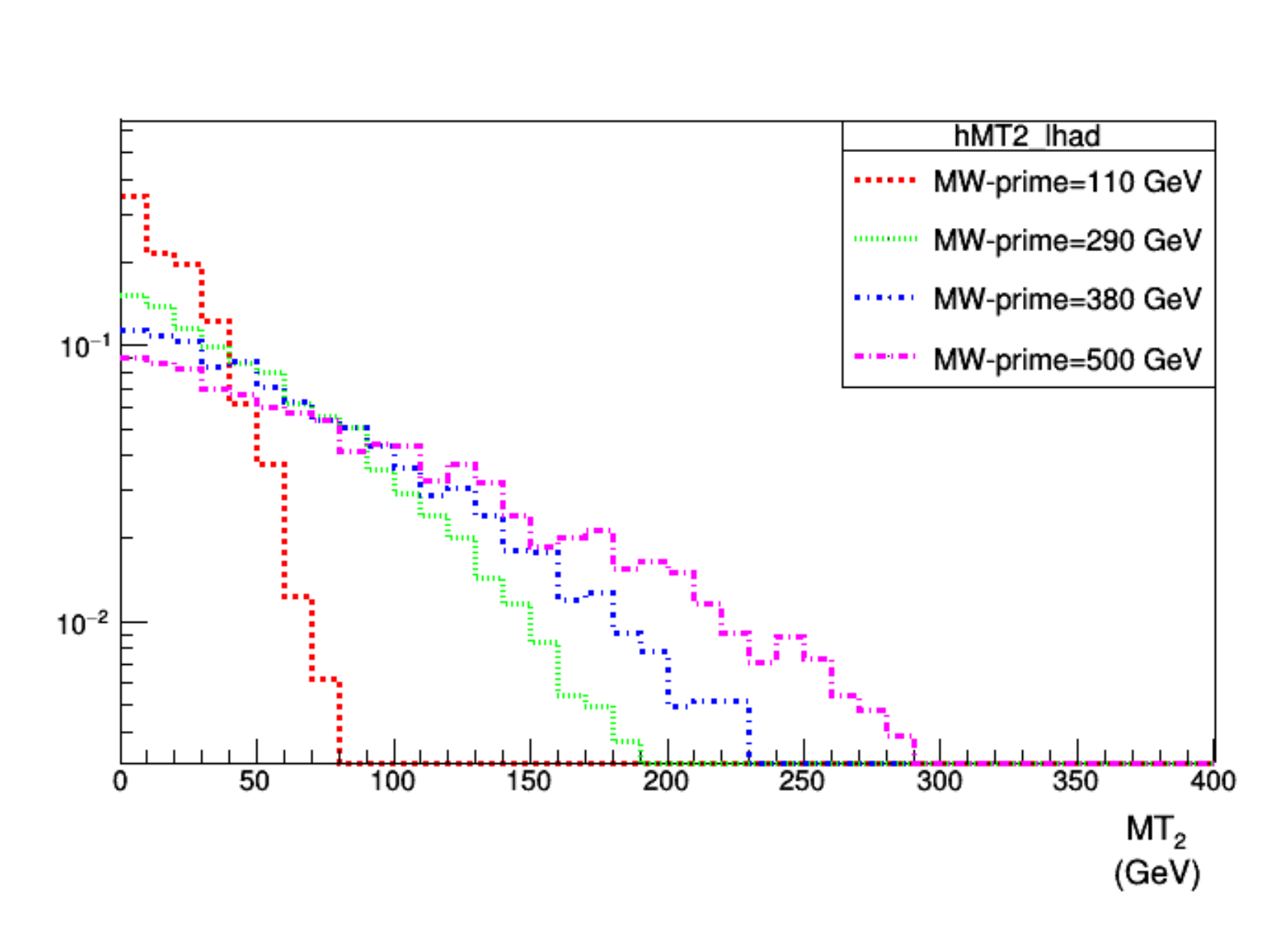} }}%
\caption{$M_{T2}$ distribution in for different mass of W' in different channels}%
\label{fig7}%
\end{figure}

\begin{figure}%
\centering
\subfloat[]{{\includegraphics[width=8cm]{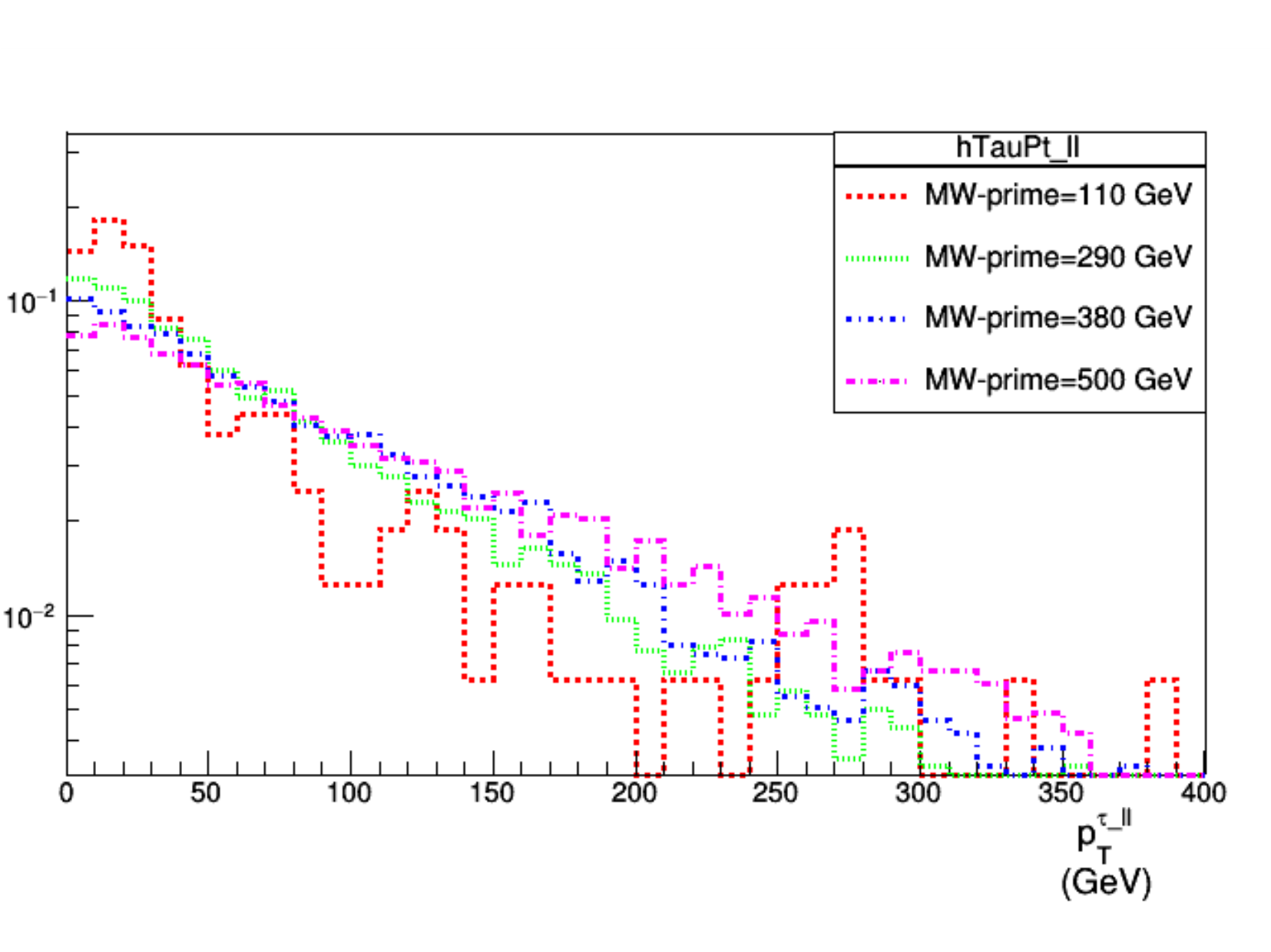} }}%
\qquad
\subfloat[]{{\includegraphics[width=8cm]{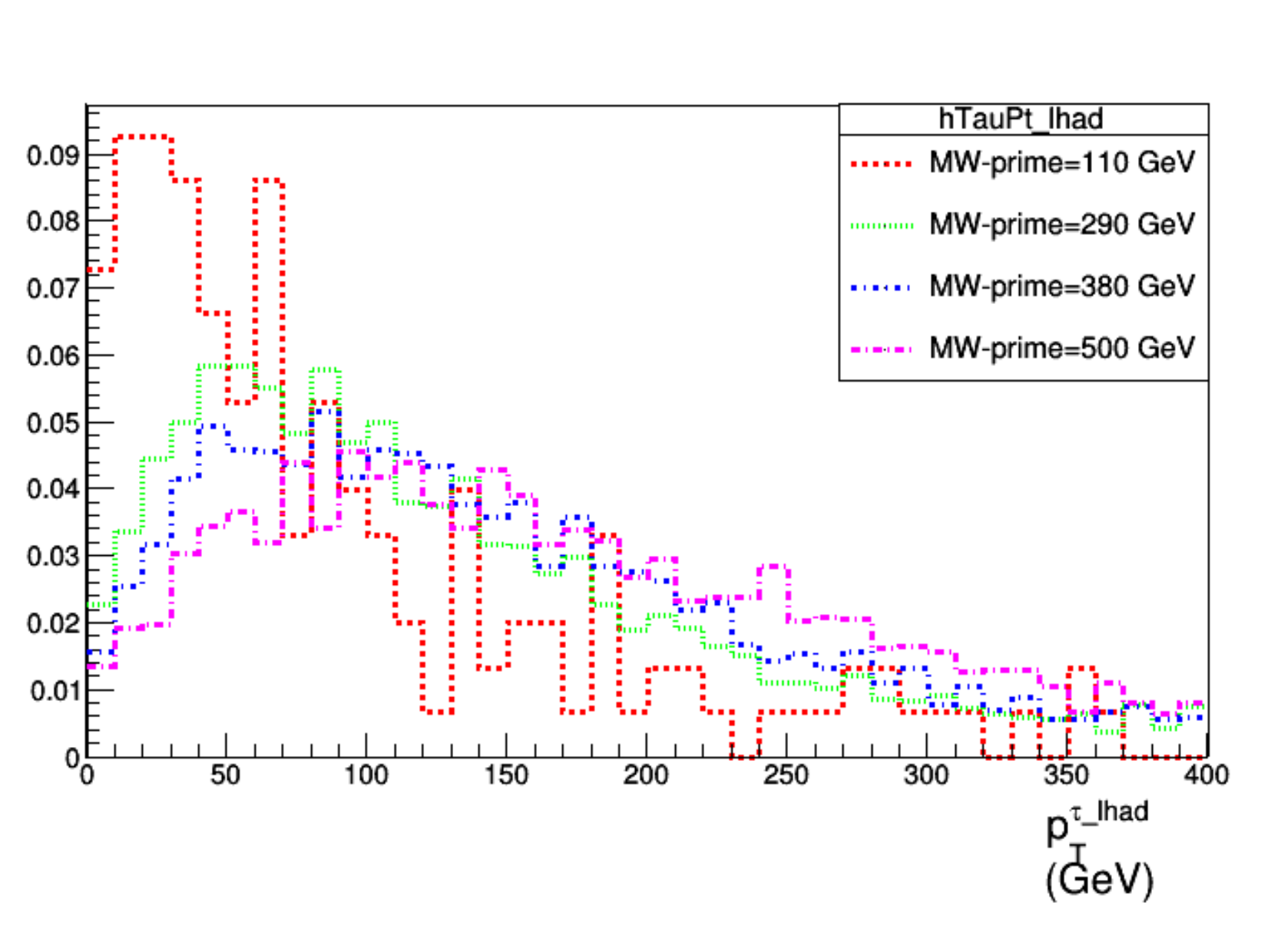} }}%
\caption{Transverse momentum distributions in different channels}%
\label{fig8}%
\end{figure}
	
\subsection{Background Events} 
	Background events are additional interactions that originate in $pp$ collision and are mainly studied in two categories. In one class gluon and quarks jets are misidentified as $\tau_h$ and in second with genuine $\tau_h$ candidate. In the first class the dominant source are the $W+$ jets and QCD multi jets and in the second case dominated events are $Z+$ jets, Higgs bosons, di-bosons and $t\bar{t}$. Background estimations are given in detail below.

\subsection{QCD Multi jets}  

 For the $e\tau_h$ and $\mu\tau_h$ decays, it has been found that the events selected with selection cut $M_{T2} < 90$ GeV is useful to discard the SM background events. But for $\tau_h \tau_h$ events, two separate SR’s are defined as events with $M_{T2} < 90$ are regarded as SR1 and events with $40 < M_{T2} < 90 $GeV and $\Sigma M^{\tau}_{T} > 250$ GeV are regarded as SR2, where $M^{\tau}_{T}$ is the sum of the transverse mass of two $\tau_h$ objects. In signal region two hadronic jets mis-identified, appeared as a pair results into QCD multi jets production. Isolation variables are used to specify genuine and misidentified $\tau_{h}$ candidate. One signals region and three control region are specified for the estimation of QCD multi jets selecting threshold on search variable $M_{T2}$ (MET) or such that $M_{T2 }$ $\mathrm{>}$ 90 GeV to 40 GeV and $\mathrm{>}$ 250 GeV to 100 GeV. One loose $\tau_{h}$ at least is selected with same sign. The non-QCD event are subtracted based on expectation of Mote Carlo (MC) simulation. The search variable are not related, isolation misidentified candidate where QCD multi jets dominated. In the two signal region SR-1 ($M_{T2\ }$$\mathrm{>}$ 90GeV) and SR-2 ($M_{T2}$ $\mathrm{<}$ 90 GeV) are estimated in the table \ref{tab7} below.

\subsection{$W+$ jets back ground }
From MC simulation, $W+$ jets are zero for remaining events in channels, while due to statistical errors in simulation sample large statistical uncertainty is there. The $W+$ background contribution from simulation are taken by formula                                                  
\begin{equation}
\label{eq6}
N_{SR} =\epsilon_{FS} \times N_{BFS}
\end{equation}
where\\
    $N_{SR }$= W+ jets in signal region \\
    $N_{BFS}$ = before final selection ($M_T2$ $\mathrm{>}$ 90 GeV for SR-1 and $\mathrm{>}$ 250 GeV for SR-2) \\
    $\epsilon_{FS} $= Final selection efficiency

The table \ref{tab8} gives the background with statistical uncertainty in two signal region for $W+$ jets.
\begin{table}[ht]
\begin{center}
\begin{tabular}{|c|c|} \hline	
Signal region & W+ jets  \\ \hline 
SR-1 & 0.70$\pm$0.21 \\ \hline 
SR-2 & 4.36$\pm$1.05 \\ \hline 
\end{tabular}
 \caption{ QCD multi jets backgrounds}
  \label{tab7}
\end{center}
\end{table}
	
\subsection{Drell-Yan backgrounds}
	This backgrounds comes from MC simulation. Different lepton pairs ($ee$, $\tau$$\tau$, $\mu$$\mu$) are included in the production. Due to misidentified probability $\tau$${}_{h}$ , contribution from $Z \to \tau \tau \to ll$ and $Z \to ll$ is very small for $l \to \tau_{h}$. For $\tau_{h} \to l$ misidentified probability is also very small, and the DY contribution to background from $Z \to \tau \tau \to \tau_{h} \tau_{h}$ and $Z \to \tau \tau \to l \tau_{h}$ are very dominant. The contribution from $Z \to \tau \tau \to \tau_{h}\tau_{h}$ very low in $l \tau_{h}$ channel. This is suitable in $\mu \tau_{h}$ control region. Table \ref{tab9} gives the estimation of DY background in $l \tau_{h}$ is given for genuine $\tau_{h}$ .
	
\begin{table}[ht]	
\begin{center}
\begin{tabular}{|c|c|c|c|} \hline	
Signal region & DY-back ground \\ \hline 
e $\tau_{h}$ & 0.19$\mathrm{\pm}$0.04 \\ \hline 
$\mu$ $\tau_{h}$ & 0.25$\mathrm{\pm}$0.06 \\ \hline 
$\tau_{h}$ $\tau_{h}$ SR-1 & 0.56$\mathrm{\pm}$0.07 \\ \hline 
$\tau_{h}$ $\tau_{h}$ SR-2 & 0.81$\mathrm{\pm}$0.56 \\ \hline 
\end{tabular}
 \caption {DY background}
 \label{tab8}
\end{center}
\end{table}
		
\subsection{Misidentified $\boldsymbol{\tau}_{h\ }$in l$\boldsymbol{\tau}_{h\ }$channels background}
	
The misidentified $\tau_{h}$ contribution in $l \tau h$ channels is assumed by a method in which probability of genuine isolated misidentified $\tau_h$ passes through tight isolation taken into account. The number of $\tau_{h}$ loose isolation candidate, when $\tau_{h}$ pass through loose isolation and signal selection is given by

\begin{equation}
\label{eq7}
N_{l} = N_g + N_m
\end{equation}                                      Where $N_{l}$ are loose, $N_{g}$ are genuine and $N_{m}$ are number of misidentified candidate. The number of tight candidate at tight selection as given by 
	
\begin{equation}
\label{eq8}
N_t = r_m (N_t- r_g N_l/r_m - r_g)    
\end{equation}  
	
Here $r_{m}$ ($r_{g}$) gives the probability for loosely selected misidentified (genuine) $\tau_{h}$ that passes through tight selection. Eliminating $N_{g}$ gives

\begin{equation}
\label{eq9}
r_{m} N_{m} = r_{m} (N_{t} - r_{g} N_{l}) / r_{m-r}
\end{equation}
	
Contamination of misidentified $\tau_{h}$ is given by the product $r_{m}$ $N_{m}$ in the signal region. The total misidentified events $l \tau_{h}$ channel are summarized in table \ref{tab10}.
\begin{table}[ht]
\begin{center}
\begin{tabular}{|c|c|c|c|} \hline	
 Signal region  & Total Misidentified \\ \hline 
 e $\tau$${}_{h}$ & 3.30$\mathrm{\pm}$3.35 \\ \hline 
 $\mu$$\tau$${}_{h}$ & 8.15$\mathrm{\pm}$4.59 \\ \hline
\end{tabular}
\caption{ Misidentified $\tau_{h}$ backgrounds}
 \label{tab9}
\end{center}
\end{table}
\noindent The combined all four signal regions background are summarized in table below including Di bosons jets (vv), $t\bar{t}$ jets (tx) and higgs bosons jets (hx).
\begin{table}[ht]
\begin{center}
\begin{tabular}{|c|c|c|c|c|} \hline
Backgrounds & \multicolumn{4}{c|}{Signal Regions} \\ \hline 
&  e $\tau$${}_{h}$ & $\mu$$\tau$${}_{h}$ & $\tau$${}_{h\ }$$\tau$${}_{h}$ SR-1 & $\tau$${}_{h\ }$$\tau$${}_{h}$ SR-2 \\ \hline 
DY & 0.19$\mathrm{\pm}$0.04 & 0.25$\mathrm{\pm}$0.06 & 0.56$\mathrm{\pm}$0.07 & 0.81$\mathrm{\pm}$0.56 \\ \hline 
vx,vv,hx & 0.03$\mathrm{\pm}$0.03 & 0.19$\mathrm{\pm}$0.09 & 0.19$\mathrm{\pm}$0.03 & 0.75$\mathrm{\pm}$0.35 \\ \hline 
W+ jets & 3.3$\mathrm{\pm}$3.35 & 8.15$\mathrm{\pm}$4.59 & 0.70$\mathrm{\pm}$0.21 & 4.36$\mathrm{\pm}$1.05 \\ \hline 
QCD multi jets &  0 & 0 & 0.13$\mathrm{\pm}$0.06 & 1.15$\mathrm{\pm}$0.39 \\ \hline 
Standard model total  & 3.52$\mathrm{\pm}$3.35 & 8.59$\mathrm{\pm}$4.59 & 1.58$\mathrm{\pm}$0.23 & 7.07$\mathrm{\pm}$1.3 \\ \hline 
Observed & 3 & 5 & 1 & 2 \\ \hline 
\end{tabular}
 \caption{ Total backgrounds events.}
 \label{tab10} 
 \end{center}
 \end{table}

\section{Exclusion}
The compatibility of the observed data with the expected signals being tested can be quantitatively performed using statistical analysis. Bayesian and frequent tests are the two most famous approaches used for this compatibility test. In this study, Bayesianic approach is used to set Limit on mass of $W^{'}$. This statistic based on Bayes theorem \cite{lab18}.
\begin{equation} 
\label{eq10}
P\left(A\vee B\right) = \frac{P\left(B\mathrel{\left|\vphantom{B A}\right.\kern-\nulldelimiterspace}A\right)P\left(A\right)}{P\left(B\right)} 
\end{equation} 
This theorem gives the conditional probability of Event $A$ when given $B$. This probability may relate to the experiment when $A$ is considered hypothesis test. In this study, $A$ is replaced with the new heavy gauge boson $W^{'}$ as a hypothesis test, while $B$ is considered as expected results. For hypothesis (Observed data) to be true $P (A/B)$ is the probability in this theorem.
To set Limits, statistical analysis is performed for which it is compulsory to choose the parameter of interest. The background events, expectation of signals and data are used to determine the probability density for this parameter. The parameter of interest selected in this study is $\sigma B$, which is also a very commonly chosen parameter in different searches of $W^{'}$ that are published. $\sigma B$ is the product of cross section ($\sigma$) of signal ($pp \to W^{'} W^{'}$) and Branching ratio of $W^{'}$ decay into the required final stat ($W^{'} \to \tau \nu \nu$). The limit for the mass exclusion may be calculated by comparing cross sections predicted upper limit by the theory.
\begin{figure}%
\centering
\subfloat[]{{\includegraphics[width=8cm]{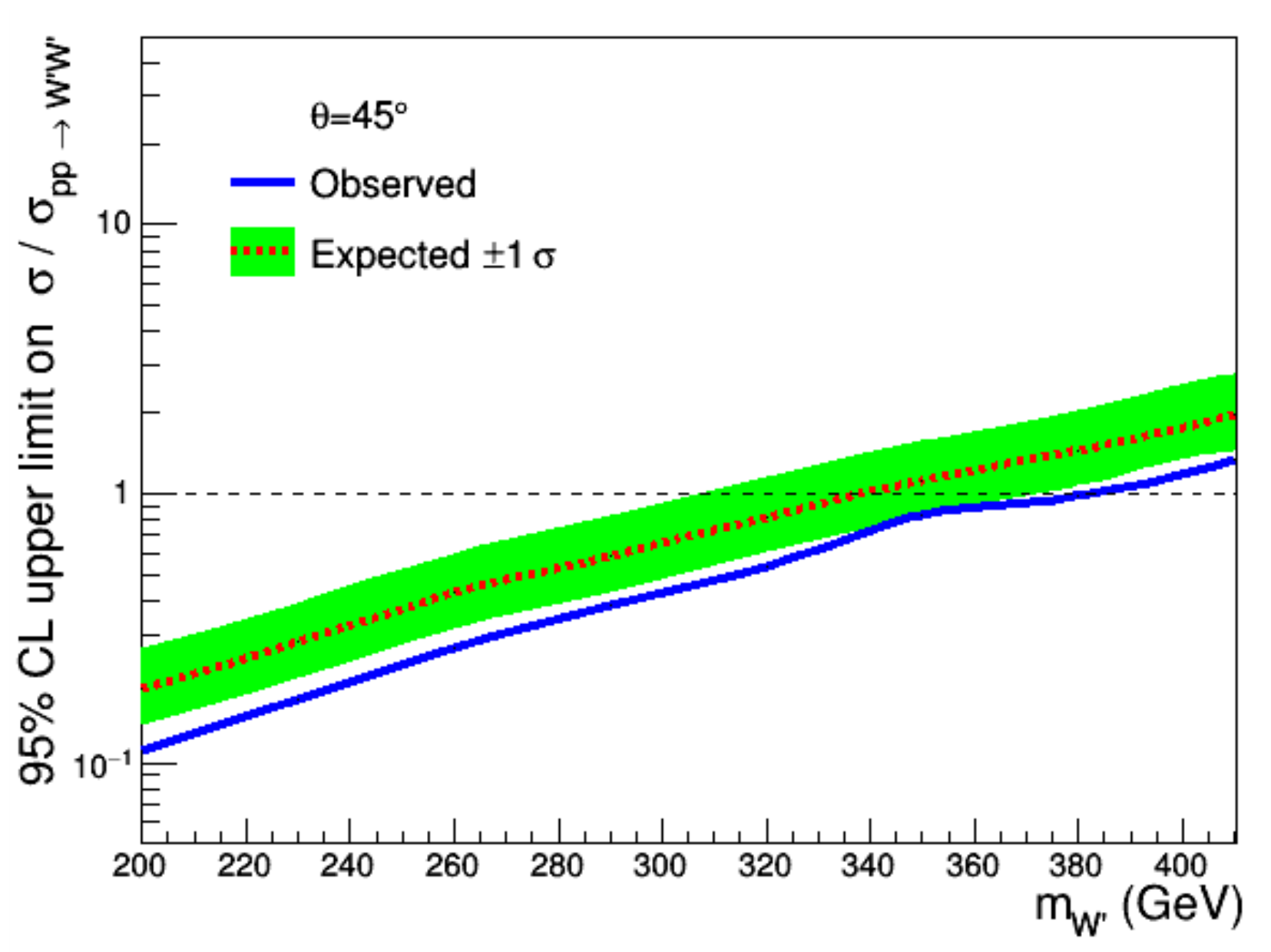} }}%
\qquad
\subfloat[]{{\includegraphics[width=8cm]{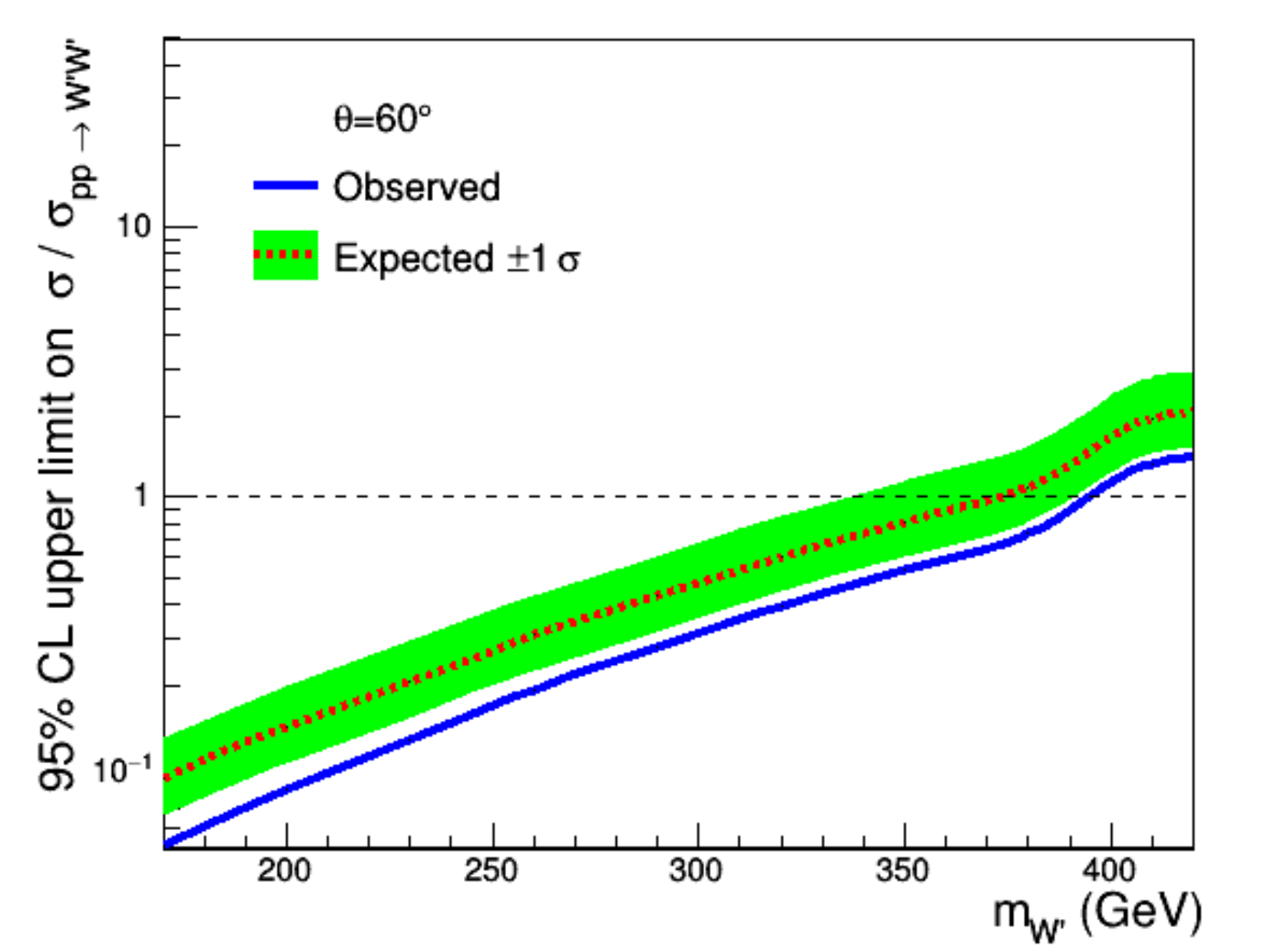} }}%
\caption{Mass at the different mass hypothesis of $W^{'}$ shown when mixing angle is 45 (left) and when mixing angle is 60 (right)}%
\label{fig9}%
\end{figure}

\begin{figure}%
\centering
\subfloat{{\includegraphics[width=8cm]{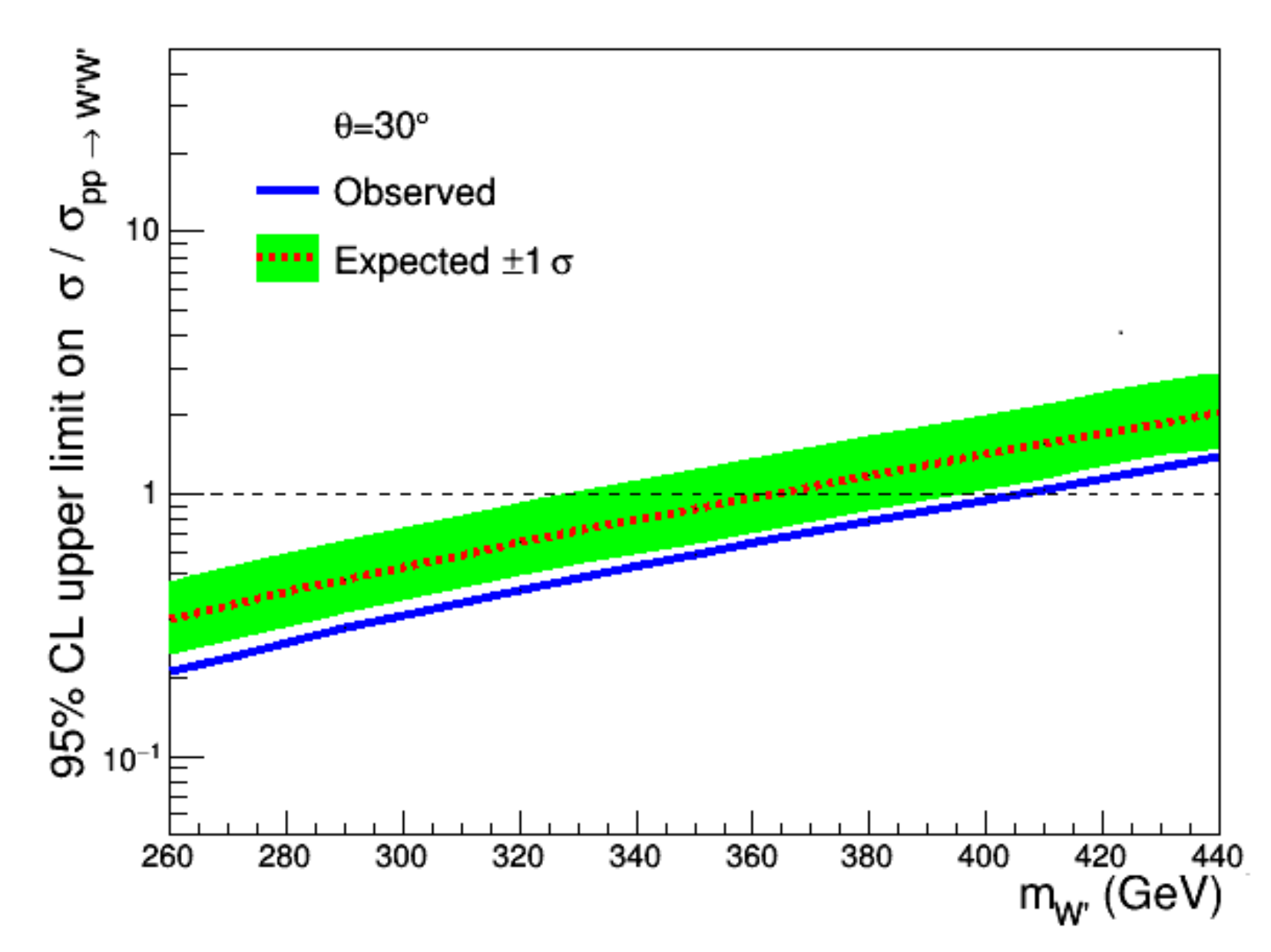} }}%
\qquad
\subfloat{{\includegraphics[width=8cm]{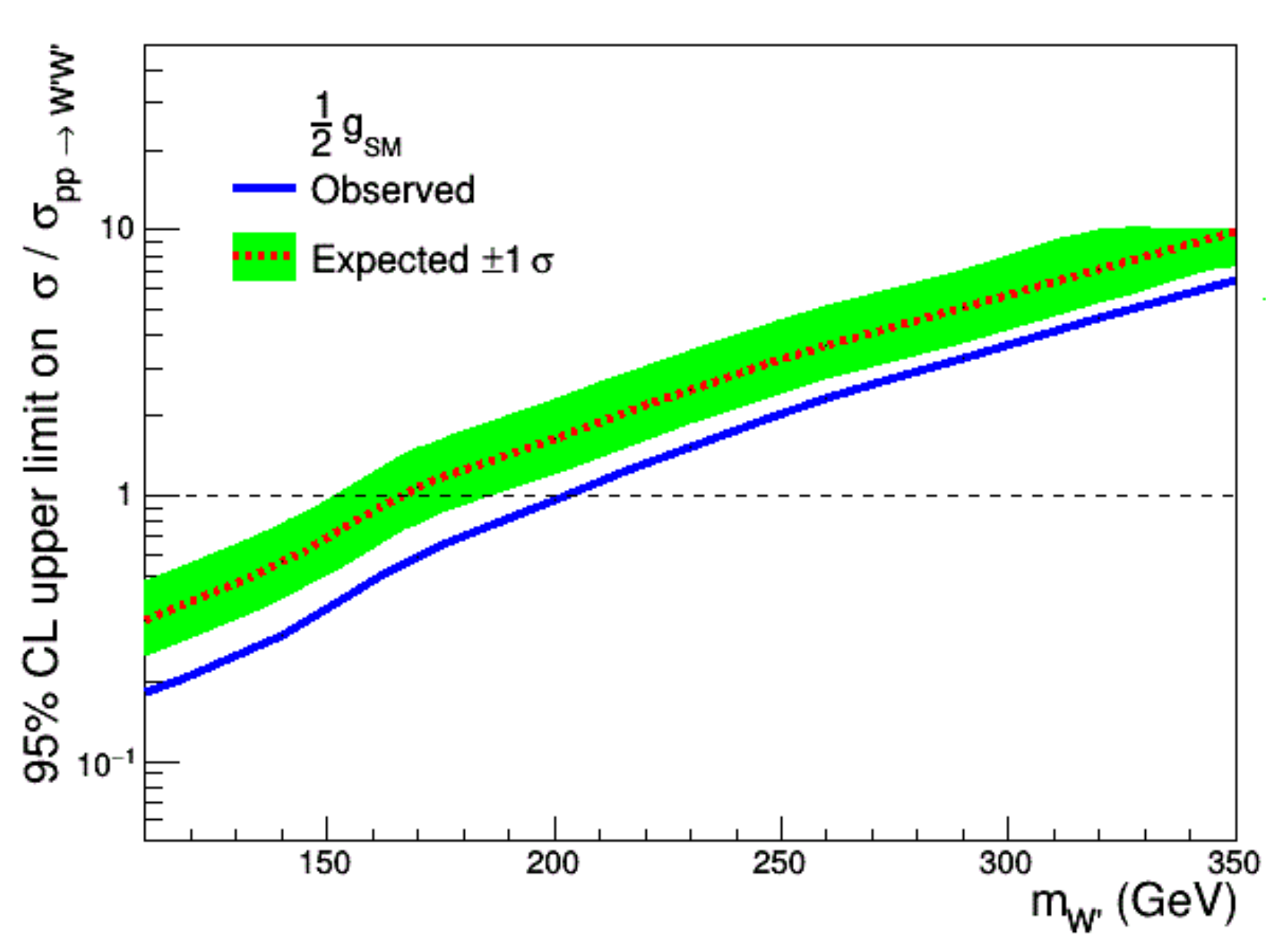} }}%
\caption{The input W' masses as a function of exclusion in 95\% Confidence Level}%
\label{fig10}%
\end{figure}

\begin{figure}%
\centering
\subfloat{{\includegraphics[width=8cm]{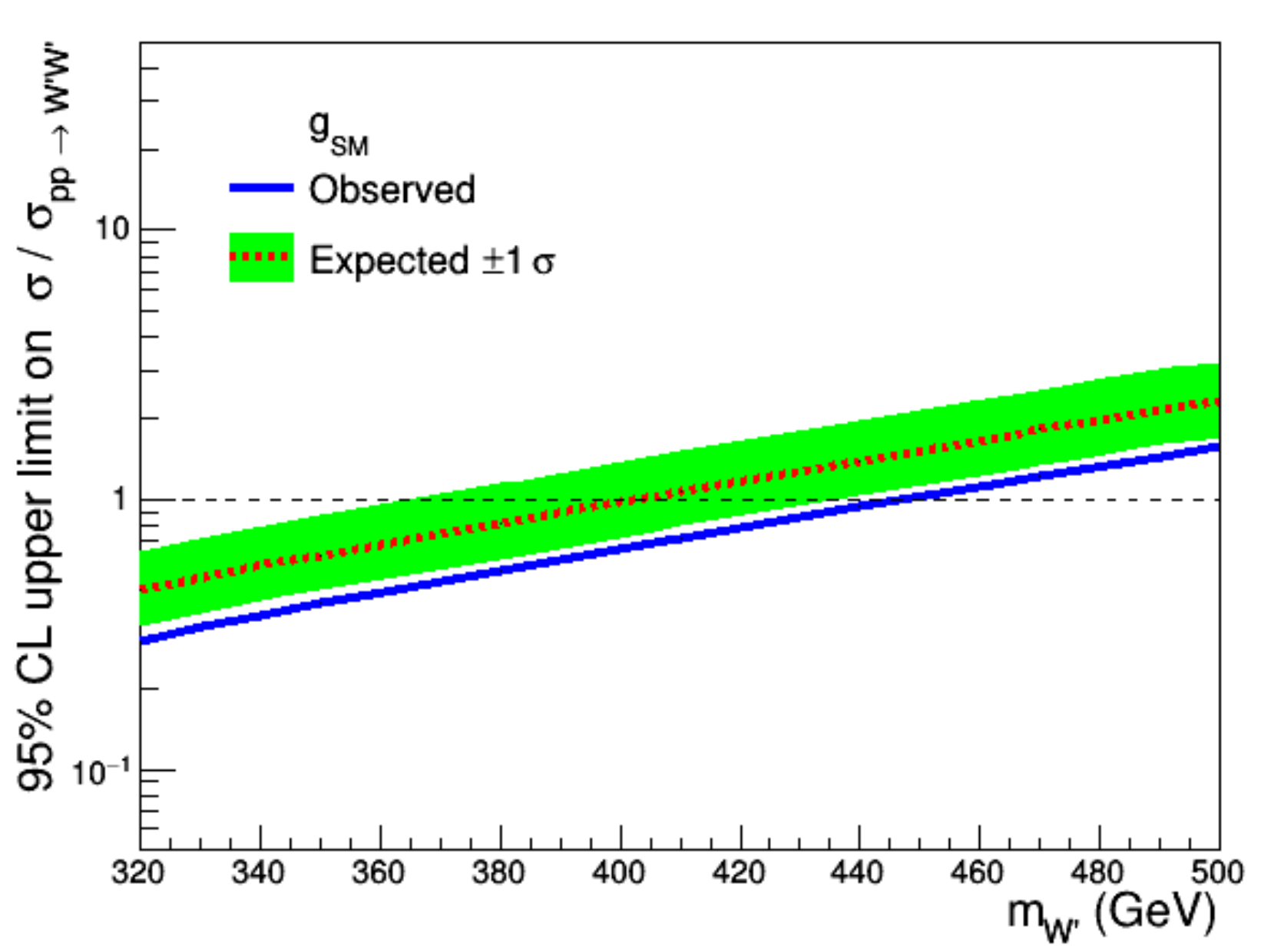} }}%
\qquad
\subfloat{{\includegraphics[width=8cm]{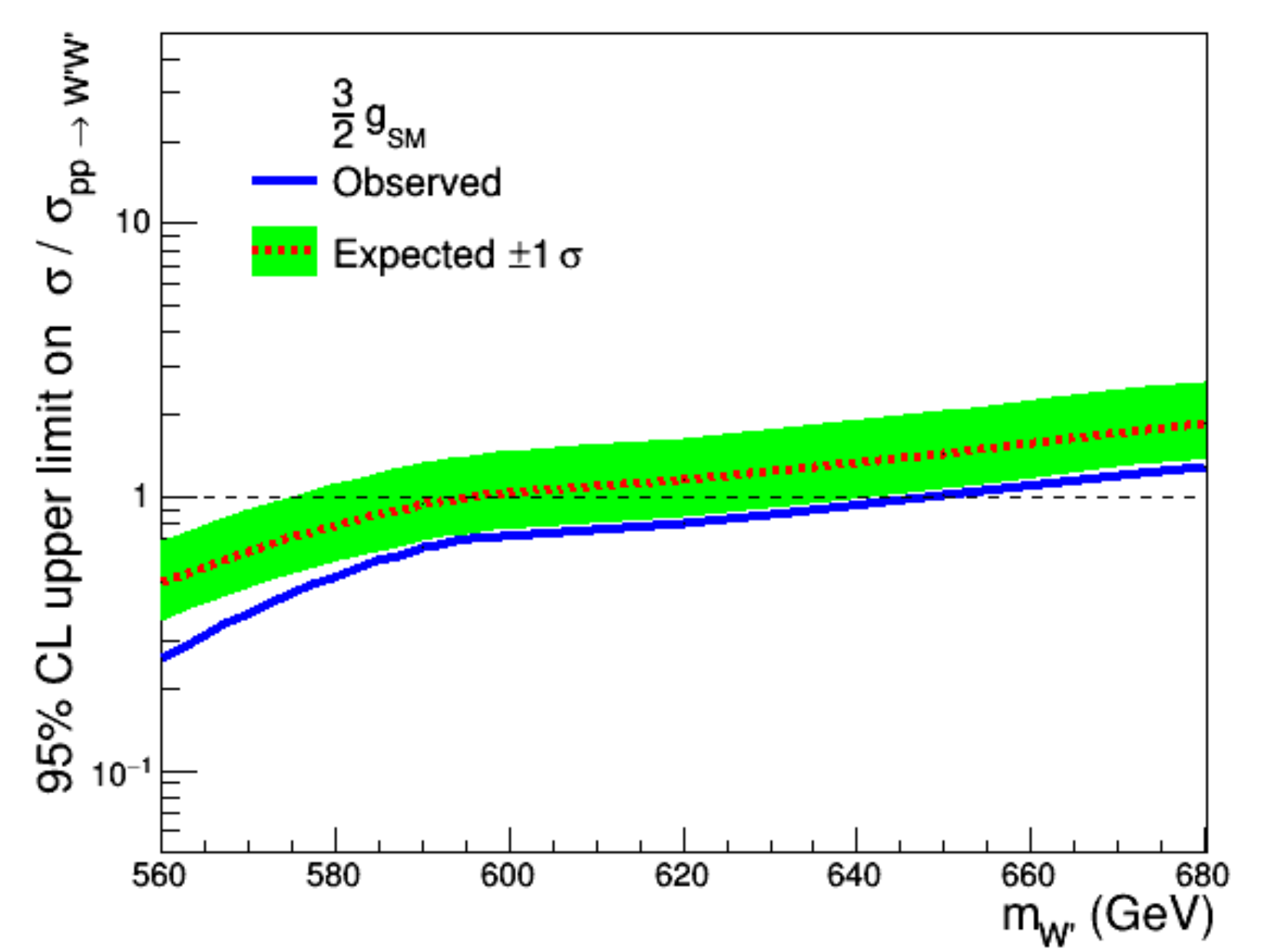} }}%
\caption{Mass of $W^{'}$ assumed for different mass hypothesis, shown observed and expected limits when coupling is exactly equal to SM coupling (left) and when coupling is one and half times the SM coupling (right)}%
\label{fig11}%
\end{figure}
Constraint on mass of $W^{'}$ can be set to exclude the lower mass at $95\%$ confidence level. The signal strength is given by the ratio $\sigma$ / $\sigma_{pp \to W^{'} W^{'}}$ which can be evaluated by applying the method of semi-bayesian ratio that is implemented in ROOT \cite{lab19}. Different results are obtained for different cross section and efficiency to set limit on mass. The standard model like and other limits are given in the table and shown in fig. \ref{fig9}. For standard model type the mass of $W^{'}$ upto 445 GeV are excluded.
This method was repeated for different scenarios and observed that limit is proportional to coupling, that the limit increases when $g'_{L}$ is increased and decreases when $g'_{L}$ is decreased. Different observed and expected limits with uncertainty of $\pm$1$\sigma$ are given in fig. \ref{fig10}, and table \ref{tab11} gives the summary of different coupling scenarios for observed and expected limits. Transverse mass distribution reconstructed from lepton pairs from each mass hypothesis assumed and is shown in fig. \ref{fig11}. As seen in table and figure that, the observed limit is always higher than the expected which could be due to the fact that large expected backgrounds in different signals region than the observed data given in the background summary table \ref{tab10}. 
We observed that compared to direct searches, the results are lower but any new model can be helpful to put possible constraints having same final state using this model, with no need of real detector response to simulate. 
\begin{table}[ht]
\begin{center}
\begin{tabular}{|c|c|c|} \hline
Mixing scenarios & Observed  & expected \\ \hline 
SM & 445 & 400 \\ \hline 
$\frac{1}{2} \times g_{SM}$  & 200 & 160 \\ \hline 
$\frac{3}{2} \times g_{SM}$ & 645 & 595 \\ \hline 
$\theta = 30$  & 405 & 365 \\ \hline 
$\theta = 45$  & 395 & 375 \\ \hline 
$\theta = 60$ & 380 & 340 \\ \hline 
\end{tabular}
\caption{ The observed and expected mass limits on mass of $W^{'}$}	
\label{tab11}
\end{center}
\end{table}

\section{Conclusion}
The $W^{'}$ pair production is easily accessible as we increase the center of mass energy for $pp$ collision. The production cross section and the decay width are calculated for different coupling strengths and for different masses. It has been observed that with the increasing mass of $W^{'}$ the decay width is increasing while the production cross section is decreasing as expected. The signal region efficiencies are found invariant with coupling strength but change with mass. This shows that it only depends upon the kinematics of the process which is related to $W^{'}$ mass, which is increasing with $W^{'}$ mass. For transverse parameters (Momentum, missing energy and mass) treatment is done which gives a distribution plot of these parameters. The distribution shows that with increasing mass of $W^{'}$ makes it accessible to get harder objects.

We have used the selection efficiencies provided for the same final state by the CMS experiment rather to fill in the complex situation of simulation for full response of detectors. These efficiencies are used for the yield of favorite signals. Statistical analysis tools are used to check the observed results with the signal yields that make it easy to set a lower limit on $W^{'}$ mass. The lower limits for the different scenarios are reported when different coupling strengths are used. For the coupling constant same as the Standard Model, it is reported that the mass below 445 GeV at the confidence level of $95\%$ are excluded. This exclusion limit may be raised up to 645 GeV when different coupling strengths are used.

\section{Acknowledgment}
We gratefully acknowledge support from the Simons Foundation and member institutions. The current submitted version of manuscript is available on arXiv pre-prints home page https://arxiv.org/pdf/2003.08558.pdf arXiv:2003.08558.




\clearpage
\end{document}